\documentclass[showpacs,aps,prd,nofootinbib,floatfix,amsmath,amssymb]{revtex4}
\usepackage{graphicx}
\usepackage{multirow}
\usepackage{epstopdf}
\usepackage{slashbox}
\usepackage{amssymb}

\makeatother

\begin{document}

\title{New physics effects in the Higgs trilinear self-coupling through one-loop radiative corrections}
\author{A. Moyotl\footnote{Present address: Ingenier\'ia en Mecatr\'onica, UPPue, Tercer Carril del Ejido "Serrano" s/n  San Mateo Cuanal\'a. Juan C. Bonilla, Puebla, Pue., M\'exico.}}
\email[E-mail:]{amoyotl@fis.cinvestav.mx}
\author{S. Chamorro-Solano\footnote{Present address: Departamento de Ciencias Naturales y Exactas, Universidad de la Costa, Calle 58 num. 55-66,
Barranquilla, Colombia.}}
\author{H. Castilla-Valdez}
\author{M. A. P\'erez}
\affiliation{Departamento de F\'isica, CINVESTAV, Apdo. Postal 14-740, 07000 M\'exico, D. F., M\'exico.}


\date{\today}

\begin{abstract}
We compute the one-loop corrections to the triple Higgs boson self-interaction $hhh$ in the framework of the Standard Model, the Two Higgs Doublet Model type III and the Littlest Higgs Model with T parity. Our results are compared with previous results for the SM. In particular, we find that an imaginary part for the $\hat{\lambda}_{hhh}$ form factor is induced when one of the Higgs boson legs is off-mass shell with 4-momentum magnitude higher than the Higgs boson mass. This contribution is sensitive to virtual effects of the Higgs boson self-interaction, that induces a radiative correction to the $hhh$ coupling of order 11\% in the on-shell scheme $\lambda_{hhh}(m_h^2,m_h^2,4m_h^2)$. However, the radiative corrections associated to the new degrees of freedom of the THDM type III and the LHM  with T-parity are smaller and comparable to the $W^\pm$ and $Z^0$ gauge bosons one-loop corrections. Accordingly, the one-loop corrected Higgs self-coupling induces minimal deviations on the SM prediction in double Higgs boson production due to gluon fusion at the LHC.
\end{abstract}


\date{\today}

\maketitle

\section{Introduction}
The study of the trilinear Higgs boson self-coupling in extensions of the Standard Model (SM) has received increased interest recently \cite{Kanemura:2015mxa,Kanemura:2002vm,Kanemura:2004ch,Arhrib:2015hoa,Ginzburg:2015iba,Kanemura:2015fra,Baglio:2016ijw,He:2016sqr,Kanemura:2016lkz,Camargo-Molina:2016moz}. In particular, a precise measurement of this self-coupling  will determine the structure of the Higgs potential and thus could confirm that the observed scalar boson with mass of 125 GeV at the LHC \cite{Aad:2012tfa,Chatrchyan:2012xdj} really corresponds to the Higgs boson predicted by the SM. It has been pointed out that the trilinear Higgs boson self-coupling could be measured in the direct determination of Higgs boson pair production in both $e^+e^-$ colliders \cite{AguilarSaavedra:2001rg,GutierrezRodriguez:2009uz} and the LHC \cite{Cao:2015oxx,Goertz:2013kp,Papaefstathiou:2012qe,Baur:2002qd,Baglio:2012np,Djouadi:1999rca,Dawson:1998py}. Furthermore, it has been found that the Higgs boson pair production could resolve the degeneracy between the anomalous couplings in single Higgs boson production \cite{Cao:2015oaa}. However, it was found that due to the uncertainties on the Higgs boson pair cross sections in both CMS and ATLAS, it is not clear if a meaningful measurement of the Higgs boson self-coupling could be possible in this case \cite{Flechl:2015foa}. On the other hand, it was found that the determination of the Higgs boson self-coupling via loop effects may be competitive with the direct determination from Higgs boson pair production in both $e^+e^-$ colliders and the LHC. This situation has been explored in the measurement of observables associated to the Higgs boson interactions \cite{Kanemura:2015mxa,Englert:2014uua} in future experiments at the high luminosity (HL)-LHC \cite{ATLAS:2013hta,CMS:2013xfa} and lepton colliders like the International Linear Collider (ILC)\cite{Baer:2013cma,Asner:2013psa}. Probing the Higgs boson self-coupling by indirect effects induced by radiative corrections has been studied also in the decay mode $h \to ZZ^* \to Z l^+l^-$ \cite{Castilla-Valdez:2015sng}, the Higgs boson interaction to gluons and photons \cite{Gorbahn:2016uoy,Hankele:2006ma}, in addition to the single Higgs boson production at the LHC \cite{Degrassi:2016wml,Bizon:2016wgr}. Although, the double Higgs production can be also evidence of new physics scenarios at LHC, such as the next-to-minimal supersymmetric Standard Model \cite{Barbieri:2013hxa}, large extra dimensions in the Arkani-Hamed-Dimopoulos-Dvalli scenario \cite{Etesami:2015caa} or the Randall-Sundrum model \cite{Zhang:2015mnh}.

In the SM, the Higgs boson self-coupling $hhh$ appears at tree level and has the form $i3 m_h^2/v$. However, at one loop level the Higgs boson self-coupling $hhh$ includes a correction given as:
\begin{eqnarray}
\Gamma_{hhh}= i\frac{3 m_h^2}{v} \lambda_{hhh},
\end{eqnarray}
where $\lambda_{hhh}$ is a form factor. On other hand, for the contribution of some SM extensions (SME), we define the deviation with respect to SM as $\Delta \lambda=\hat{\lambda}_{hhh}^\text{SME}/\hat{\lambda}_{hhh}^\text{SM}$. In this sense, deviations with respect to this prediction have been observed in the effective Lagrangian approach \cite{Kanemura:2015mxa,Kanemura:2002vm,Kanemura:2004ch}, in radiative one-loop effects induced in the Two Higgs Doublet Models \cite{Kanemura:2002vm,Kanemura:2004ch,Ginzburg:2015iba,Hankele:2006ma}, in the Minimal Supersymmetric Model with a light stop \cite{Kanemura:2004ch,Kanemura:2015fra}, by a heavy neutrino \cite{Baglio:2016ijw}, and finally in the real \cite{He:2016sqr,Kanemura:2016lkz,Camargo-Molina:2016moz} and the complex singlet \cite{Camargo-Molina:2016moz} extensions of the SM.

It is convenient to point out that the $hhh$ vertex has not on-shell direct mapping to an S-matrix element, and as a consequence other one-loop diagrams should be considered in order to get a full physical transition amplitude for a given process. In the present paper we are interested in determining, in a complete renormalization scheme, the one-loop radiative corrections to the $h^*hh$ vertex with two Higgs bosons on-shell mass and the other one off-shell mass. This vertex is appropriated in the study of the production of a pair of Higgs bosons at the LHC. In this case, it has been known for a long time \cite{Glover:1987nx} that the one-loop box diagram involved in the process $gg \to hh$ has a very important effect that cannot be neglected since there is a cancelation against the contribution with a virtual Higgs boson $gg \to h^* \to hh$ by an order of magnitude. In particular, Kanemura et al. \cite{Kanemura:2016lkz} have found recently that a large deviation of the one-loop corrected $hhh$ coupling may induce a sizeable deviation of the cross section for the process $gg \to hh$. Since our results for the one-loop correct $hhh$ coupling are of order 11\%, we will not address in the present paper the complete calculation involved in the pair production of Higgs bosons at hadron colliders.

Our results for the radiative corrections to the $h^*hh$ vertex can be easily generalized to the vertex $hh^*h^*$ with two virtual Higgs bosons which is involved in the calculation of the complete physical process for the decay $h \to ZZ^* \to Zl^+l^-$ \cite{Castilla-Valdez:2015sng}. In the present paper we are interested in extending the one-loop radiative corrections to the $h^*hh$ vertex by including flavor-changing neutral couplings (FCNC) of the Higgs boson in the framework of Two Higgs Doublet Model (THDM) type III, and mirror fermions in the Littlest Higgs Model (LHM) with T-parity, respectively. We have found that the FCNC corrections to the Higgs boson self-coupling may be greater than the tau lepton or the bottom quark contributions, while the mirror fermions effects are close to the contribution induced by the SM gauge bosons in the Higgs in the on-shell scheme $\lambda(m_h^2,m_h^2,4m_h^2)$. On other hand, we found that in the Higgs on-shell scheme, the virtual effect of the Higgs boson self interaction induces a radiative correction of the same order of magnitude than the top-quark virtual effect in the SM, which in turn produces an overall radiative correction to the $hhh$ coupling of order of $11$\%. However, there are several studies that have found sizable deviations on the SM $hhh$ vertex induced by new degrees  of freedom heavier than the top-quark mass and these effects could be large enough as to be detected in the process $gg \to hh$ \cite{Kanemura:2016lkz,Baglio:2017kqy}.

The plan of the paper is the following. In section 2 we review the theoretical set-up of THDM type III and LHM with T-parity, necessary for our analysis. In section 3 we present all analytical expressions for the self-coupling $hhh$ at one-loop level in the SM framework, the quark contributions in THDM type III and the mirror fermions contributions in LHM with T-parity. The numerical results are presented in section 4 and finally, concluding remarks and outlook are presented in section 5. Details of the renormalization scheme used in our calculation are included in Appendices A and B.

\section{Theoretical framework}
In this section we present the relevant aspects of the Higgs sector of the Two Higgs Doublet Model type III and the Littlest Higgs Model with T-parity, which will be needed to obtain the one-loop correction to the self-coupling $hhh$. For this purpose, only the Higgs boson interaction to the respective fermions is analyzed.
\subsection{Two Higgs Doublet Model type III}
In the Two Higgs Doublet Model type III, the flavor changing processes are associated with the charged ($H^\pm$) and the neutral scalar bosons ($h^0,H^0,A^0$) at tree level. These interactions are described in detail in \cite{HernandezSanchez:2012eg}; we are interested in the interaction of the Higgs boson and quarks type up given by:

\begin{eqnarray}
{\mathcal L}_{\bar{u}_iu_jh^0}=-\frac{g}{2m_W} \bar{u}_i\bigg[ m_{u_i} \xi_h^u\delta_{ij}-\frac{(\xi_H^u+\xi_h^u\cot \beta)}{\sqrt{1+\cot^2 \beta}}\frac{\sqrt{m_{u_i}m_{u_j}}}{\sqrt{2}} \tilde{\chi}_{ij}^u \bigg]u_j h^0 \label{lagrangian-2HDM-III}
\end{eqnarray}
with $h^0$ the light neutral Higgs boson of the THDM, $\xi_h^u=\cos \alpha/\sin \beta$, $\xi_H^u=\sin \alpha/\sin \beta$ and $\tilde{\chi}^u$ is a complex mixing matrix which takes a specific form according to the texture used for the Yukawa matrices \cite{GomezBock:2005hc}. In particular, the SM with just flavor-conserving neutral couplings is recovered with $\sin(\beta-\alpha)=1$ \cite{GomezBock:2005hc}. Since we are interested in reproducing the SM results already measured for the Higgs boson at the LHC, it is convenient to stay with $\sin(\beta-\alpha)\simeq1$ and we will introduce a new parameter defined as $\chi=\pi/2-(\beta-\alpha)$ \cite{Kanemura:2015mxa}. We will assume also that the dominant FCNC is given by the $h^0tc$ coupling.

\subsection{Littlest Higgs Model with T-parity}
The LHM is an effective theory based in the collective symmetry breaking approach, where in the first stage the global group $SU(5)$ is broken to $SO(5)$ at a scale $f$ in the TeV range, via the symmetric tensor of vacuum expectation:
\begin{eqnarray}
\Sigma_0=\left( \begin{array}{ccc}
\mathbf{0}_{2\times2} & \mathbf{0}_{2\times1} & \mathbf{1}_{2\times2} \\
\mathbf{0}_{1\times2} & 1 & \mathbf{0}_{1\times2} \\
\mathbf{1}_{2\times2} & \mathbf{0}_{2\times1} & \mathbf{0}_{2\times2} \end{array} \right).
\end{eqnarray}
Simultaneously, the gauged subgroup  $[SU(2)_1\times U(1)_1]\times[SU(2)_2\times U(1)_2]$ of $SU(5)$ is broken to the electroweak SM group $SU(2)_L\times U(1)_Y$. Finally, the gauged group $SU(2)\times U(1)$ is broken to $U(1)_\text{em}$ via the usual Higgs mechanism; however, this Higgs potential corresponds to the Coleman-Weinberg potential which is generated by one-loop radiative corrections. On other hand, from the global symmetry breaking of $SU(5)/SO(5)$, we generated 14 Nambu-Golstone bosons and four of them were absorbed by heavy gauge bosons ($W_H^{\pm}$, $Z_H$, $A_H$); the remaining ten Nambu-Golstone bosons are parametrized by the nonlinear sigma model
\begin{eqnarray}
{\mathcal L}_\Sigma=\frac{f^2}{8}\text{Tr}|{\mathcal D}_\mu\Sigma|^2,
\end{eqnarray}
where $\Sigma=e^{i \Pi/f} \Sigma_0e^{i \Pi^T/f}$, while the field $\Pi$ and the covariant derivative ${\mathcal D}_\mu$ are given in \cite{Blanke:2007db,Reuter:2013iya}. The implementation of T-parity on the gauge fields consists in exchanging the two $SU(2)\times U(1)$ factors; consequently the gauge coupling of the two $SU(2)\times U(1)$ factors are equal and therefore the number of free parameters is reduced. The T-parity in the fermion sector is introduced by implementing two doublets $SU(2)_1$ and $SU(2)_2$ such that under T-parity the even linear combination is associated to the SM $SU(2)$ doublet, while the T-odd combination is associated with the so called mirror fermions. These fermions acquire mass through $SU(5)$ and a $T$ invariant Yukawa interaction:
\begin{eqnarray}
{\mathcal L}_\text{mirror}^\text{LHM+T}=-\kappa_{ij} f\big( \bar{\Psi}_2^i \xi+ \bar{\Psi}_1^i \Sigma_0\Omega\xi^\dag\Omega\big)\Psi_R^j\label{lagrangian-LHM+T};
\end{eqnarray}
here $\kappa_{ij}$ is a mixing matrix and, in principle it is different for each mirror fermion, $\xi=e^{i \Pi/f}$, $\Omega=\text{diag(1,1,1-1,1,1)}$ and $\Psi_{1,2,R}$ is a multiplet with five components. Then, after expanding the Lagrangian (\ref{lagrangian-LHM+T}) at ${\mathcal O}(v^2/f^2)$, the masses of the respective mirror fermions are given by:
\begin{eqnarray}
m_{d_H}&=&m_{\ell_H}=\sqrt{2}\kappa_{ii}f\label{mdh},
                                       \\
m_{u_H}&=&m_{\nu_H}=\sqrt{2}\kappa_{ii}f\Big( 1-\frac{v^2}{8f^2}\Big)\label{muh},
\end{eqnarray}
where the mirror down-quark (mirror charged lepton) receives only corrections of order ${\mathcal O}(v^3/f^3)$. Moreover, the Higgs couplings to the mirror up-quark and mirror heavy neutrinos are given by
\begin{eqnarray}
h \bar{u}_Hu_H=h \bar{\nu}_H\nu_H\sim\frac{i\kappa_{ii}}{2\sqrt{2}}\frac{v}{f} \label{huH-uH}.
\end{eqnarray}
On other hand, the Higgs boson does not have direct couplings to mirror down-quarks and mirror charged leptons.

\section{One-loop corrections to the self-coupling $hhh$}

In this section we present the different contributions to the $\lambda_{hhh}$ form factor induced at one-loop level. For this purpose, we take the SM contributions as reference to compare our results obtained for THDM type III and LHM with parity T. In order
to compute the general one-loop vertex corrections of the self-coupling $hhh$ with off-shell scalar bosons and to satisfy the Bose symmetry, we need additional vertices with the different permutations of the particles 4-momenta. However, in the final expressions we will have only one off-shell scalar boson  i.e. $h(m_h^2)h(m_h^2)h^*(q^2)$. This structure of the $\lambda_{hhh}$ form factor allows to analyze a Higgs boson pair production from a third off-shell Higgs in hadronic or $e^+e^-$ collisions, with known 4-momenta. In this sense, the Higgs self-coupling $hhh$ was analyzed in processes such as $gg$ double-Higgs fusion ($gg \to hh$), $VV$ double-Higgs fusion ($qq' \to hh qq'$, $V=W^\pm,Z$), double Higgs-strahlung ($q\bar{q}'\to Vhh$), associated production with top-quarks $q\bar{q}/gg\to t\bar{t}hh$ \cite{Baglio:2012np} and high energy photon-photon collisions via the $\gamma\gamma \to t\bar{t}hh$ process \cite{GutierrezRodriguez:2011gi}. Nevertheless, in these analyses the self-coupling $hhh$ is independent of the 4-momenta of any Higgs boson. Finally, we have used the Feynman parametrization to resolve the tensorial integrals and we found that the results are UV-divergent, which need to be renormalized. In order to renormalize the self-coupling $hhh$ at one-loop level  we need the counterterm contributions $\delta \lambda_{hhh}$, which cancel the respective UV-divergence and lead to a finite result:
\begin{equation}
\hat{\lambda}_{hhh}=\lambda_{hhh}+\delta\lambda_{hhh},
\end{equation}
where the counterterm of the self-coupling is built by redefinitions of parameters and the Higgs boson field as follow \cite{Bohm:1986rj,Denner:1991kt}:
\begin{equation}
\delta\lambda_{hhh}=\delta Z_e-\frac{\delta s_W}{s_W}+\frac{\delta m_h^2}{m_h^2}+\frac{e}{2s_W}\frac{\delta t}{m_Wm_h^2}-\frac{1}{2}\frac{\delta m_W^2}{m_W^2}+\frac{3}{2}\delta Z_h.\label{deltalambda}
\end{equation}
The details of these procedures are given in the Appendix A. In next section we show the general form of the different contributions to the self-coupling $hhh$, where $\Delta=2/\epsilon-\gamma+\log(4\pi)$ represents the respective UV-divergent term and $\Lambda$ is an energy cutoff.

\subsection{SM framework}
In the SM the Higgs self-coupling $hhh$ at one-loop level is induced by fermions, gauge bosons and Higgs self-coupling, and the main contribution is induced by a top-quark loop. In this subsection we will include the different corrections to self-coupling $hhh$ at one-loop level in SM.

\subsubsection{Fermionic contributions}
These contributions are induced only by three-point loops of fermions, and thus the correction to the self-coupling $hhh$ can be written as follows:

\begin{eqnarray}
\lambda_f^\text{SM}(q^2)=\frac{g^2m_f^4N_c}{3!48\pi^2 m_h^2m_W^2} \Bigg[ 18+\frac{3}{2}\Delta-\int_{x=0}^1\int_{y=0}^{1-x}\text{d}x\text{d}y\bigg[\Xi_f^{SM}(x,y,s_f,s_q)+144s_f\sum_{k=1}^3\log\Big( \frac{m_h^2}{\Lambda^2}\mu_k^2\Big)\bigg]\Bigg],\label{l-SM2}
\end{eqnarray}
where $3!$ is the number of possible permutations in the 4-momenta of the external Higgs bosons and $N_c$ is the number of color, with $N_c=3$ for quarks and $N_c=1$ for leptons. Further, $s_k^2=m_k^2/m_h^2$, $s_q^2=q^2/m_h^2$, $\mu_k^2$ can be obtained from the Feynman parameters $M_k^2$ given in appendix B, and $\Xi_f^{SM}(x,y,s_f,s_q)$ is a dimensionless function given by:

\begin{eqnarray}
\Xi_f^{SM}(x,y,s_f,s_q)&=&\big[4s_f^2+s_q^2(1-2x-2y)+2(x+y-1)\big]\frac{1}{\mu_1^2}
                                       \nonumber\\
{}&{}&+\big[4s_f^2-2x+s_q^2(2x-1)\big]\frac{1}{\mu_2^2}+\big[4s_f^2-2y+s_q^2(2y-1)\big]\frac{1}{\mu_3^2}.\label{Xi-f}
\end{eqnarray}

In particular, when we consider the Higgs on-shell scheme, we have the following simplified expression:
\begin{eqnarray}
\Xi_f^{SM}(x,y,s_f)=\frac{3(4s_f^2-1)}{s_f^2+x^2+(x+y)(y-1)}.
\end{eqnarray}

\subsubsection{Gauge bosons contributions}
In this sector there are contributions induced by the gauge bosons $V=Z^0$, $W^\pm$, which are generated by a three-point (labeled by $V_I$) and a two-point loop (labeled by $V_{II}$) diagrams of gauge bosons. Then, after solving the tensorial integral we found the following results:
\begin{eqnarray}
\lambda_{V_I}^\text{SM}(q^2)&=&\frac{g^2m_W}{3!96 \pi^2\eta^6 m_hs_V^3} \int_{x=0}^{1}\int_{y=0}^{1-x}\Xi_V^\text{SM}(x,y,s_V,s_q) \text{d}x \text{d}y, \label{lambda-VI}
                                       \\
\lambda_{V_{II}}^\text{SM}(q^2)&=&-\frac{g^2m_W}{3!1440 \pi^2 \eta^4m_hs_V^3} \big[7(2+s_q^4)-30(2+s_q^2)s_V^2\big] \label{lambda-VII},
\end{eqnarray}
where $\eta=1$ for $W^\pm$ and $\eta=\cos\theta_W$ for $Z^0$. Further, $\Xi_V^\text{SM}(x,y,s_V,s_q)$ is another dimensionless function with a large mathematical structure but the specific form of this function is not included in this report. Then, the total correction of this sector has the form $\lambda_{W+Z}^\text{SM}=\lambda_{Z_I}^\text{SM}+\lambda_{Z_{II}}^\text{SM}+\lambda_{W_I}^\text{SM}+\lambda_{W_{II}}^\text{SM}$.

\subsubsection{Higgs self-coupling contributions}
Finally, there is a correction induced by the Higgs self-coupling contribution, which is also generated by a three-point (labeled by $h_I$) and a two-point (labeled by $h_{I}$) loop diagrams of Higgs bosons. This correction has not been included in previous calculations in the framework of the SM \cite{Kanemura:2002vm,Kanemura:2016lkz}. We will find that its contribution is lower than the top-quark loop, but of the same order of magnitude. It is important to mention that the correction induced by the three-point loop $\lambda_{h_I}^\text{SM}$ is free of UV-divergences, and thus it was not necessary to renormalize it with the respective counterterm contribution. On other hand, we found that the contribution of the two-point loop diagram $\lambda_{h_{II}}^\text{SM}$ vanishes when the result is renormalized by the respective counterterm contribution. Then, the results are the following:
\begin{eqnarray}
\lambda_{h_I}^\text{SM}(q^2)&=& \frac{9g^2m_h^2}{3!32 \pi^2 m_W^2} \int_{x=0}^1 \int_{y=0}^{1-x} \Xi_h^\text{SM}(x,y,s_q) \text{d}x \text{d}y,\label{lambda_hI}\\
\lambda_{h_{II}}^\text{SM}(q^2)&=&\frac{3g^2m_h^2}{3!32\pi^2m_W^2}\bigg[ 3\Delta-2\int_{x=0}^1 \log\Big[\frac{m_h^2}{\Lambda^2}\big[x(x-1)+1\big]\Big]\text{d}x+\int_{x=0}^1\log\Big[\frac{m_h^2}{\Lambda^2}\big[x(x-1)s_q^2+1\big]\Big]\text{d}x \bigg]\label{lambda_hII},
\end{eqnarray}
where the dimensionless function $\Xi_h^\text{SM}(x,y,s_q)$ is given by:
\begin{eqnarray}
\Xi_h^\text{SM}(x,y,s_q)&=&\frac{1}{1+(y-1)y+s_q^2x(x+y-1)}+\frac{1}{1+(x-1)x+s_q^2y(x+y-1)}
                                        \nonumber\\
{}&{}&+\frac{1}{1-sq^2xy+(x+y-1)(x+y)}.
\end{eqnarray}
and the dimensionless function in the Higgs on-shell scheme have the following simplified expression:
\begin{eqnarray}
\Xi_h^\text{SM}(x,y)=\frac{3}{x^2+(y-1)(x+y)+1}.
\end{eqnarray}

\subsection{THDM type III framework}
In this subsection we present the quark contributions of the THDM type III, which arise from the diagonal and nondiagonal Higgs interactions given by Eq. (\ref{lagrangian-2HDM-III}). First, the diagonal part is proportional to the SM fermion contribution, while the nondiagonal part is a new contribution due to the correction of the self-coupling $hhh$ of the THDM type III. The general case corresponds to a three-points loop with three effective vertices; but we will consider for simplicity the radiative corrections induced with only one or two effective vertices.

\subsubsection{Flavor conserving case}
The diagonal Higgs interactions given by Eq. (\ref{lagrangian-2HDM-III}) induce a small correction to the SM coupling $h^0 \bar{u}_i u_i$. We found that the correction to the self-coupling $hhh$ is proportional to fermionic contribution of the SM, i.e. $\lambda^\text{THDM-III}=\tilde{h}_{ii}^n \lambda_u^\text{SM}$, where $n=1,2,3$ represents the number of effective vertices in the loop and

\begin{eqnarray}
\tilde{h}_{ii}=\xi_h^u-\frac{\xi_H^u+\xi_h^u\cot \beta}{\sqrt{2}\sqrt{1+\cot^2\beta}}\tilde{\chi}_{ii}^u.
\label{h-eff}
\end{eqnarray}

\subsubsection{Flavor changing case}

In the case of nondiagonal Higgs interactions, we used only two effective vertices $h^0 \bar{u}_i u_j$ and after solving the tensorial integral, we obtained the following result:

\begin{eqnarray}
\lambda_\text{FC}^\text{THDM-III}(q^2)&=& \frac{g^2N_c}{96\pi^2}\frac{m_h^2}{m_W^2}\frac{(\xi_H^u-\xi_h^u\cot \beta)^2}{(1+\cot^2 \beta)}\frac{\text{Re}(\tilde{\chi}_{ij}^u\tilde{\chi}_{ji}^{u*})}{(2)3!}\times
                                        \nonumber\\
{}&{}&\bigg{\lbrace} 12s_i s_j(s_i^2+s_i s_j+s_j^2)-\Bigg[ s_i^2s_j \int_{x=0}^1\int_{y=0}^{1-x} \Xi(x,y,s_i,s_j,s_q)\text{d}x\text{d}y
                                        \nonumber\\
{}&{}&+24(2s_i+s_j)\bigg[\frac{3}{2}\Delta-\sum_{k=1}^{3}\int_{x=0}^1\int_{y=0}^{1-x}\log\Big(\frac{m_h^2}{\Lambda^2}M_k^2\Big)\text{d}x\text{d}y\bigg] +(i\rightleftarrows j)\Bigg]\bigg{\rbrace},\label{l-eff2}
\end{eqnarray}
where $2\text{Re}(\tilde{\chi}_{ij}^u\tilde{\chi}_{ji}^{u*})=\tilde{\chi}_{ij}^u\tilde{\chi}_{ji}^{u*}+\tilde{\chi}_{ij}^{u*}\tilde{\chi}_{ji}^{u}$, the expressions for the $M_k^2$ functions are included in the Appendix B, while the symbol $i\rightleftarrows j$ represents the exchange the quark type in the loop and, therefore the factor $(2)3!$ is introduced to obtain the average of these contributions. Furthermore, we introduce the following dimensionless function:

\begin{eqnarray}
\Xi(x,y,s_i,s_j,s_q)&=& \Big[s_i(4s_i^2-s_q^2)(x+y)-s_j[2s_j^2+4s_js_i-(2s_i^2-s_q^2+2)](x+y-1)\Big]\frac{1}{M_1^2}
                                        \nonumber\\
{}&{}&+\Big[s_i[4s_i^2(x+y)-2x+s_q^2(x-y)]+s_j[2s_i^2-4s_is_j-(2s_j^2-s_q^2)](x+y-1)\Big]\frac{1}{M_2^2}
                                        \nonumber\\
{}&{}&+\Big[s_i[4s_i^2(x+y)-2y-s_q^2(x-y)]+s_j[2s_i^2-4s_is_j-(2s_j^2-s_q^2)](x+y-1)\Big]\frac{1}{M_3^2}.
\end{eqnarray}

 Note that the one-loop correction $\lambda_\text{FC}^\text{THDM-III}$ is symmetrical to the exchange of masses in the loop ($s_i\rightleftarrows s_j$). Finally, it is important to mention that with appropriate changes of masses and coupling constants, equation (\ref{l-eff2}) reproduces the result of the SM given by Eq. (\ref{l-SM2}).

\subsection{LHM with T-parity framework}
In analogy with the SM, in the LHM with T-parity there are contributions arising from heavy mirror fermions, heavy gauge bosons and heavy scalars bosons. These contributions are suppressed by the energy scale $f$\footnote{The masses of heavy bosons and heavy scalar are directly proportional to the symmetry breaking scale $f$, as well as the mirror fermion masses \cite{Han:2003wu}.}, but since the contributions of the SM fermions are proportional to the fourth power of their masses, we expect that the main contributions arise from mirror fermions in the LHM with T-parity. Therefore, we will consider only the contributions of mirror fermions. Then, from the masses (\ref{muh}), the Higgs couplings to the mirror fermions (\ref{huH-uH}) and the fermionic contributions (\ref{l-SM2}), the respective contribution is given by:

\begin{eqnarray}
\lambda_{f_H}^\text{LHM+T}(q^2)&=&\sum_{i=1}^3\frac{\kappa_{ii}^2 m_{f_H} m_W N_c}{3\pi^22^{\frac{11}{2}}3! gm_h^2}\frac{v^3}{f^3}\Bigg[ 18+\frac{3}{2}\Delta
                                        \nonumber\\
{}&{}&-\int_{x=0}^1\int_{y=0}^{1-x}\text{d}x\text{d}y\bigg[\Xi_{f_H}^{SM}(x,y,s_{f_H},s_q)+144s_{f_H}\sum_{k=1}^3\log\Big( \frac{m_h^2}{\Lambda^2}\mu_k^2\Big)\bigg]\Bigg],\label{l-LHM+T}
\end{eqnarray}
where $f_H$ can be a mirror heavy up-quark or mirror heavy neutrino; furthermore the sum comprises three families of mirror fermions. While, the $\Xi_{f_H}^{SM}$ and $\mu_k^2$ functions are the same as given in Eq. (\ref{l-SM2}) with $f\to f_H$.

\section{Numerical results and discussion}

It is important to emphasize that in the SM the self-coupling $hhh$ at tree level is independent of the 4-momentum of any scalar boson. However, at one-loop level and with at least one off-shell scalar boson, this coupling may contain real and imaginary parts. At this stage, the respective $\hat{\lambda}_{hhh}$ form factor depends on loop particles masses; if these particles are heavier than the 4-momentum of the off-shell Higgs boson, then the induced form factor is a complex function of $q^2$. Consequently, the $\hat{\lambda}_{hhh}$ form factor in the Higgs on-shell scheme is a real function for $||q||=m_h<2 m$, with $m$ the particle mass coupled to the off-shell Higgs boson. This uncoupling behavior was obtained also in the Higgs-boson form factor effects in $t\bar t$ production \cite{Gounaris:2016cyc}, the trilinear neutral gauge boson couplings $ZZZ^*$, $ZZ\gamma^*$ and $Z^*Z\gamma$ \cite{Moyotl:2015bia}, as well as in the electromagnetic \cite{Moyotl:2011yv} and weak static properties of tau lepton \cite{Bolanos:2013tda,Moyotl:2012zz}. For these reasons, we focused on the $\hat{\lambda}_{hhh}$ form factor contribution with the 4-momentum of the off-shell Higgs boson higher than the Higgs boson mass. Details of the renormalization procedure for the dominant heavy fermion contributions are included in Appendix A. Moreover, our numerical analysis has been performed without approximations and we have used $m_h=125$ GeV for the Higgs boson mass.

\subsection{SM framework}

\subsubsection{Top quark contribution}
\begin{figure}[!hbt]
\centering
\includegraphics[scale=0.95]{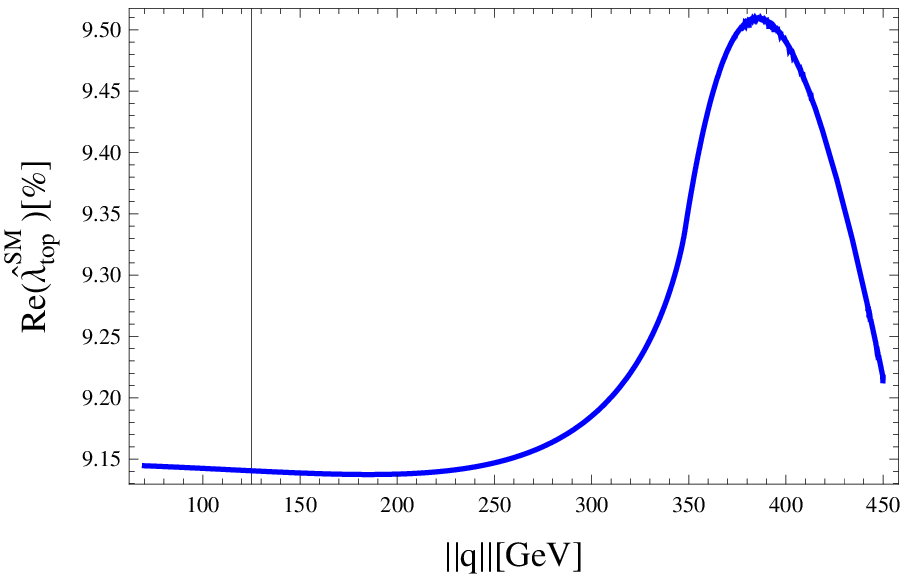}
\includegraphics[scale=0.95]{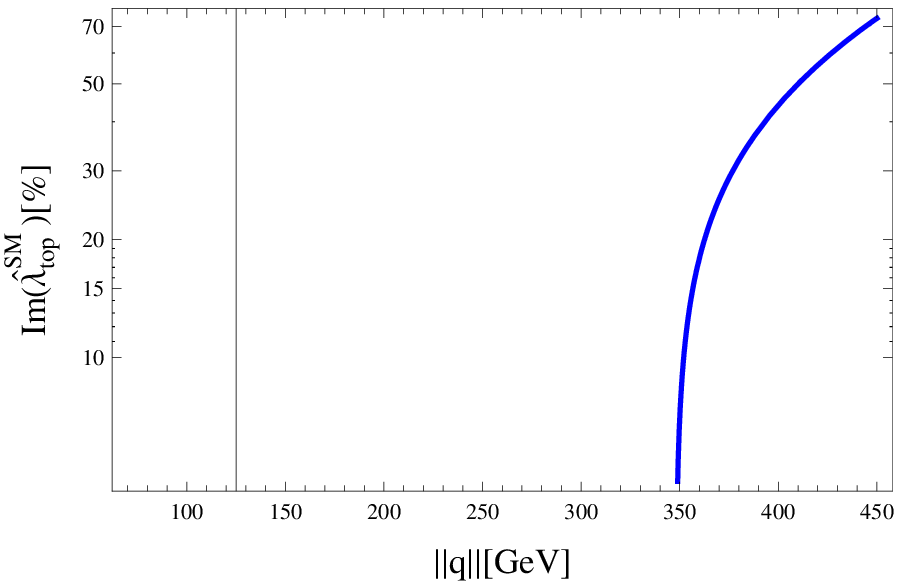}
\caption{Real (left) and imaginary (right) contributions of top-quark to the $\hat{\lambda}_\text{top}^\text{SM}$ form factor, as function of the 4-momentum magnitude of the off-shell scalar boson. There is no induced imaginary part for $||q||<2m_t$ and $\hat{\lambda}_\text{top}^\text{SM}\simeq9.14049$ \% in the Higgs on-shell scheme. The vertical lines indicate the Higgs on-shell scheme ($||q||=125$ GeV).}
\label{lambda-sm-top}
\end{figure}

We analyzed the Higgs on-shell scheme and we found $\hat{\lambda}_\text{top}^\text{SM}=9.14049 \%$ for the top-quark contribution; however, the bottom-quark ($\hat{\lambda}_\text{b}^\text{SM} \simeq 3.442\times10^{-6}\%+i6.4810\times10^{-9}\%$) and tau lepton ($\hat{\lambda}_\tau^\text{SM} \simeq 2.535\times10^{-8}\%+i7.962\times10^{-12}\%$) contributions are very suppressed. Then, in the Higgs on-shell scheme the main contribution to the $\hat{\lambda}_{hhh}$ form factor comes from the top-quark. In the following analysis we consider only the top-quark contribution. In order to compare our results with previous work, the Eq. (1) of \cite{Kanemura:2002vm} gives $\hat{\lambda}_{hhh}^\text{eff}(\text{SM})\simeq 9.8221\%$ and this result was obtained by the diagrammatic approach. On other hand, $\hat{\lambda}_\text{top}^\text{SM}\simeq 9.14693\%$ can be extracted from $\Delta \hat{\Gamma}_H^{3}$ of Eq. (32) in the reference \cite{Hollik:2001px}. These contributions are very similar to our results, but it is important to mention that they were obtained with some approximations for the 4-momentum magnitude of the off-shell scalar boson. On other hand, Figure \ref{lambda-sm-top} shows the real and imaginary parts of the $\hat{\lambda}_\text{top}^\text{SM}$ form factor as function of the 4-momentum magnitude of the off-shell scalar boson.

We obtained that there is no imaginary part for $||q||<2 m_t$. Nevertheless, we will show that the imaginary part increases considerably for higher values of $||q||$ and the real part remains basically stable, although it has a small maximum after $||q||=2 m_t$ and this numerical behavior is consistent with the analysis of \cite{Arhrib:2015hoa}. It is important to mention that a large contribution to the $\hat{\lambda}_\text{top}^\text{SM}$ form factor (for higher values of $||q||$) can violate unitarity, but we will show below that the gauge bosons and the Higgs self-coupling can reduce this large imaginary contribution.

\begin{figure}[!hbt]
\centering
\includegraphics[scale=0.95]{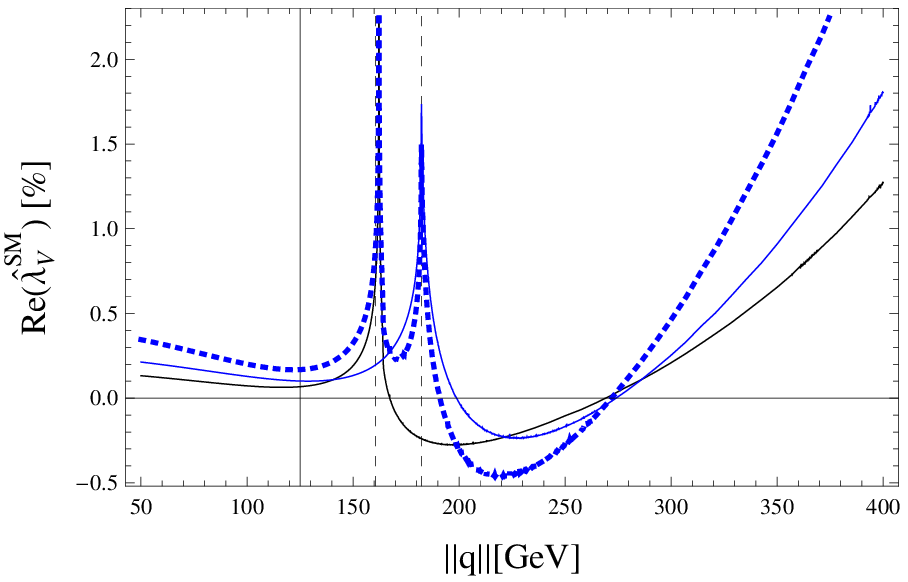}
\includegraphics[scale=0.95]{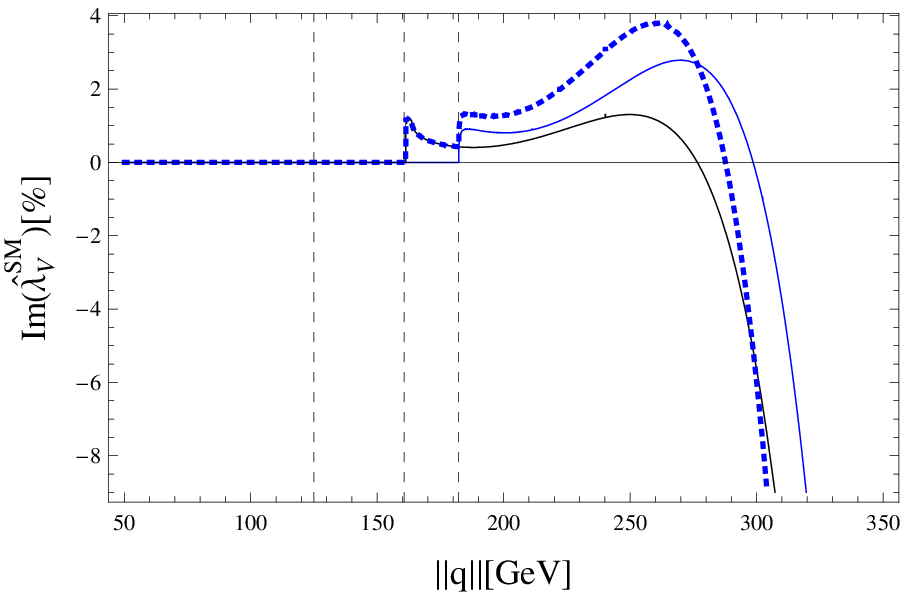}
\caption{Real (left) and imaginary (right) contributions of gauge bosons to the $\hat{\lambda}_{V}^\text{SM}$ form factor; we show the $W^\pm$ boson (black line) and $Z^0$ boson (blue line) contribution, as well as the sum of both contributions (dotted blue line) as function of the 4-momentum  magnitude of the off-shell scalar boson. There are no induced imaginary parts for $||q||<2m_V$, while in the Higgs on-shell scheme we found $\hat{\lambda}_W^\text{SM}\simeq 0.0388\%$, $\hat{\lambda}_Z^\text{SM}\simeq 0.03384 \%$ and $\hat{\lambda}_{W+Z}^\text{SM}\simeq 0.0726\%$. The solid vertical line indicates the Higgs on-shell scheme, while the dashed vertical lines corresponds to $2 m_V$.}
\label{lambda-sm-V}
\end{figure}
\subsubsection{Gauge bosons contributions}
The total contribution of this sector includes contributions from $W^\pm$ and $Z^0$ bosons; we show these contributions for each gauge boson in Figure \ref{lambda-sm-V}. There is no imaginary part for $||q||<2m_V$ again and there is a great negative contribution for higher $||q||$. For the real part, we obtain peak contributions in $||q||\simeq 2 m_V$, which agrees with the uncoupling behavior. In these regions we have corrections of order $3.524\%$ ($||q||\simeq 2 m_W$) and $1.192\%$ ($||q||\simeq 2 m_Z$), respectively; these values are in agreement with the analysis of \cite{Arhrib:2015hoa}. These regions give the greatest contributions to the gauge sector. Moreover, the real part increases smoothly with increasing $||q||$, but the self-coupling $hhh$ contribution will reduce this large contribution.

Of particular importance is the Higgs on-shell scheme, where we found $\hat{\lambda}_W^\text{SM}\simeq 0.0388\%$ and $\hat{\lambda}_Z^\text{SM}\simeq 0.03384 \%$ for the $W^\pm$ and $Z^0$ gauge bosons contributions, respectively. Then the total contribution is given by $\hat{\lambda}_{W+Z}^\text{SM}\simeq 0.0726\%$, which is approximately three orders of magnitude smaller than the top quark contribution.

\subsubsection{Higgs self-coupling contributions}
\begin{figure}[!hbt]
\centering
\includegraphics[scale=0.95]{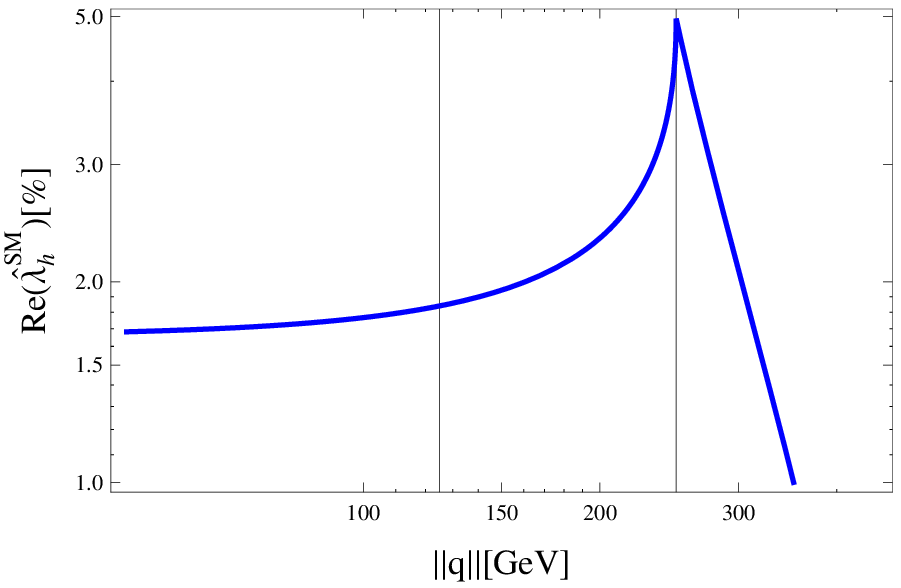}
\includegraphics[scale=0.95]{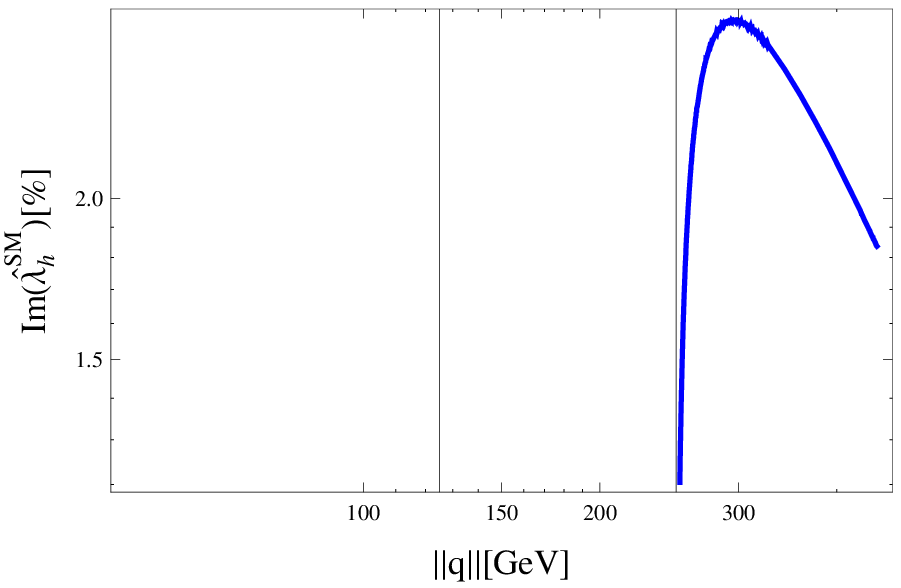}
\caption{Real (left) and imaginary (right) contributions of the Higgs self-coupling to the $\hat{\lambda}_h^\text{SM}$ form factor as function of the 4-momentum  magnitude of the off-shell scalar boson. There is no induced imaginary part for $||q||<2m_h$ and $\hat{\lambda}_h^\text{SM}\simeq1.83974$ \% in the Higgs on-shell scheme. The vertical line indicates the Higgs on-shell scheme ($||q||=125$ GeV).}
\label{lambda-sm-higgs}
\end{figure}

The contribution of this sector is shown in the Figure \ref{lambda-sm-higgs}, with similar properties  that have characterized the previous sectors. In particular for $||q||\simeq 2 m_h$ we found a correction of approximately 4.9168\% and there is a negative real contribution for higher $||q||$, which can reduce the large real part of the gauge bosons contributions. Furthermore, for higher $||q||$ there is also a negative imaginary contribution which in turn reduces the large imaginary top-quark contribution. For the Higgs on-shell scheme, we found $\hat{\lambda}_h^\text{SM}\simeq1.83974$ \%, which is a bit smaller than the top-quark contribution, but it is approximately two orders of magnitude bigger than the gauge bosons contribution. Therefore, the total SM self-coupling $hhh$ corrections at one-loop level is $\hat{\lambda}_{hhh}^\text{SM}=\hat{\lambda}_\text{top}^\text{SM}+\hat{\lambda}_{W+Z}^\text{SM}+\lambda_h^\text{SM}\simeq 11.0528\%$, which is a relativity high correction.

\subsection{THDM type III framework}
\subsubsection{Flavor conserving case}
\begin{figure}[!hbt]
\centering
\includegraphics[scale=0.6]{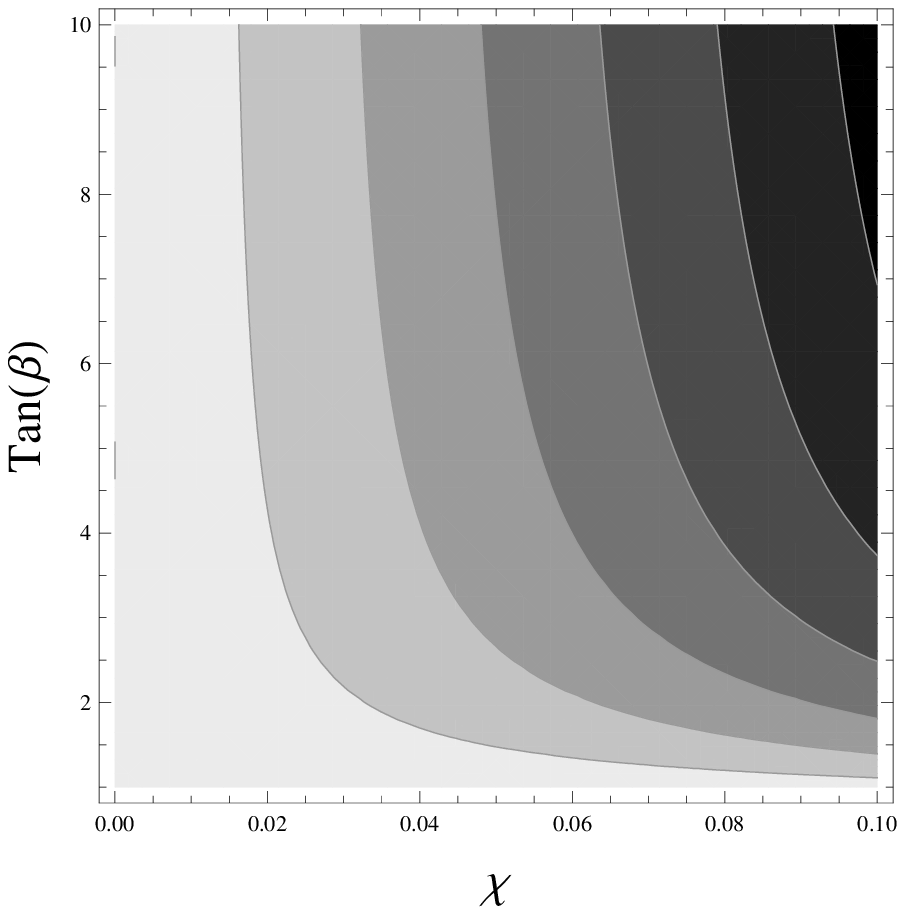}
\includegraphics[scale=1.15]{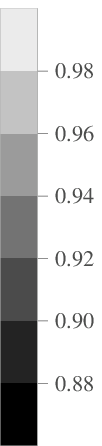}
\caption{Desviation with respect to the SM $\Delta \lambda^\text{THDM-III}=\tilde{h}_{ii}^2$ for the THDM type III in the flavor conserving case. The respective contribution is very similar to the SM result when $\chi$ is very suppressed, while it decreases when $\tan \beta$ increases.}
\label{lDelta-lambda1}
\end{figure}

As we mentioned above, in this case the radiative correction is proportional to the fermionic SM contribution and the main contribution is provided by the top-quark loop; the deviation with respect to the SM is $\Delta\lambda^\text{THDM-III}=\hat{\lambda}_\text{top}^\text{THDM-III}/\hat{\lambda}_\text{top}^\text{SM}=\tilde{h}_{ii}^n$. In the following analysis, we focus exclusively in the situation for $n=2$ and $\tilde{\chi}_{ii}=1$; the results are shown in Figure \ref{lDelta-lambda1}. It is important to note the region for small values of the $\chi$ parameter: where for smallest $\chi$, the result is basically independent of the value of $\tan\beta$. This situation leads to the SM model correction, which is consistent with the discussion presented in section II.A. On other hand, for unsuppressed values of $\chi$, the contribution decreases considerably with increasing $\tan\beta$. The larger contributions to the $\hat{\lambda}^\text{THDM-III}$ factor come from small values of $\tan\beta$.

\subsubsection{Flavor changing case}

\begin{figure}[!hbt]
\centering
\includegraphics[scale=0.95]{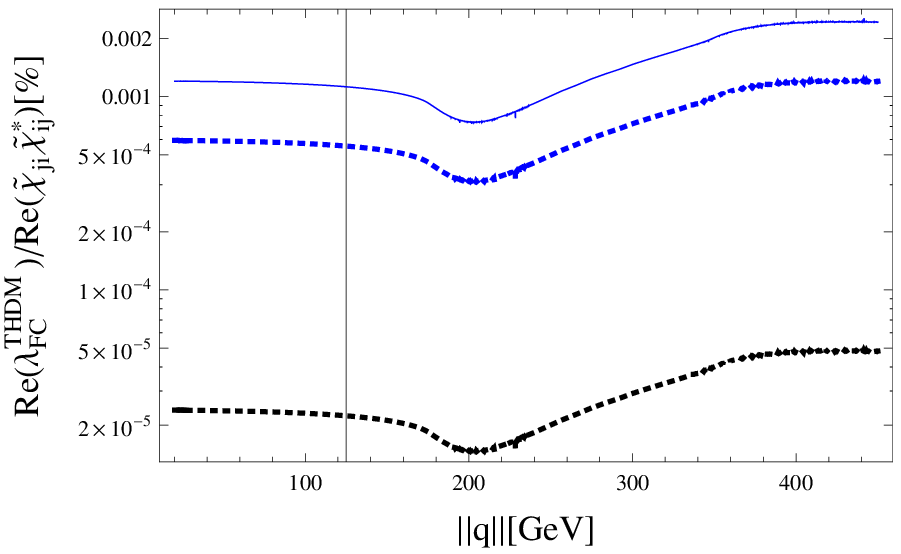}
\includegraphics[scale=0.95]{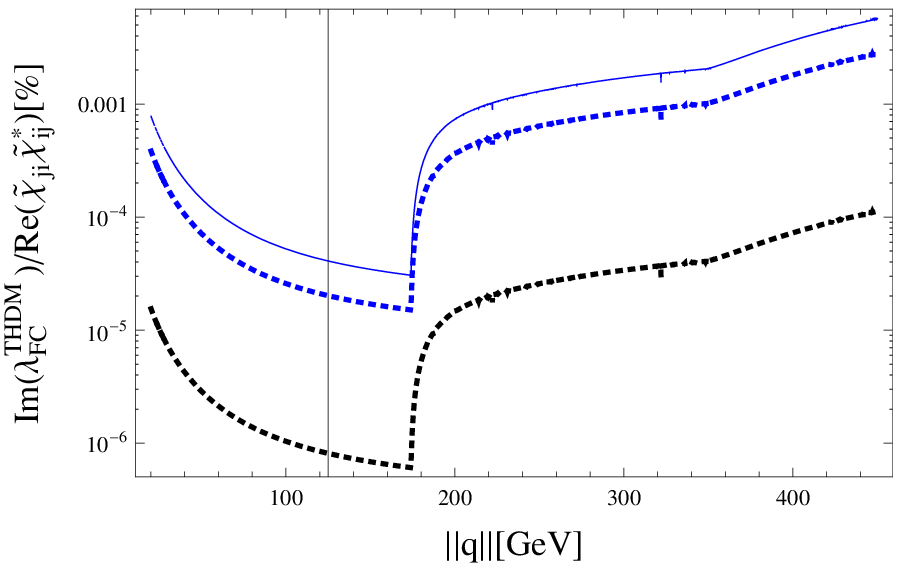}
\caption{Real (left) and imaginary (right) contribuctions of $\hat{\lambda}_\text{FC}^\text{THDM-III}$ form factor, as function of the 4-momentum magnitude of the off-shell scalar boson. We have used $\tan \beta=1$ with $|\chi|=0.20$ (blue line), $0.14$ (blue dotted line) and $0.028$ (black dotted line). The vertical line indicate the Higgs on-shell scheme ($||q||=125$ GeV).}
\label{lambda-FV-top-tan1}
\end{figure}

\begin{figure}[!hbt]
\centering
\includegraphics[scale=0.95]{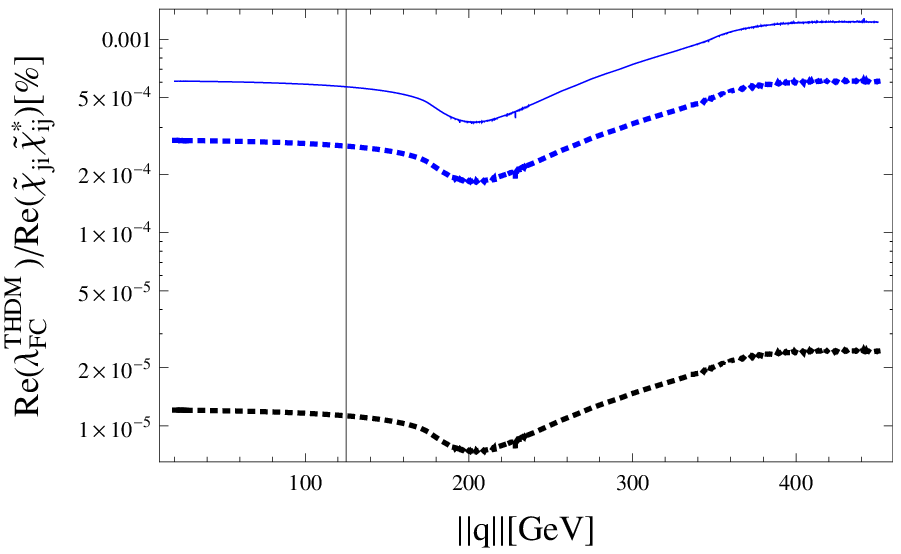}
\includegraphics[scale=0.95]{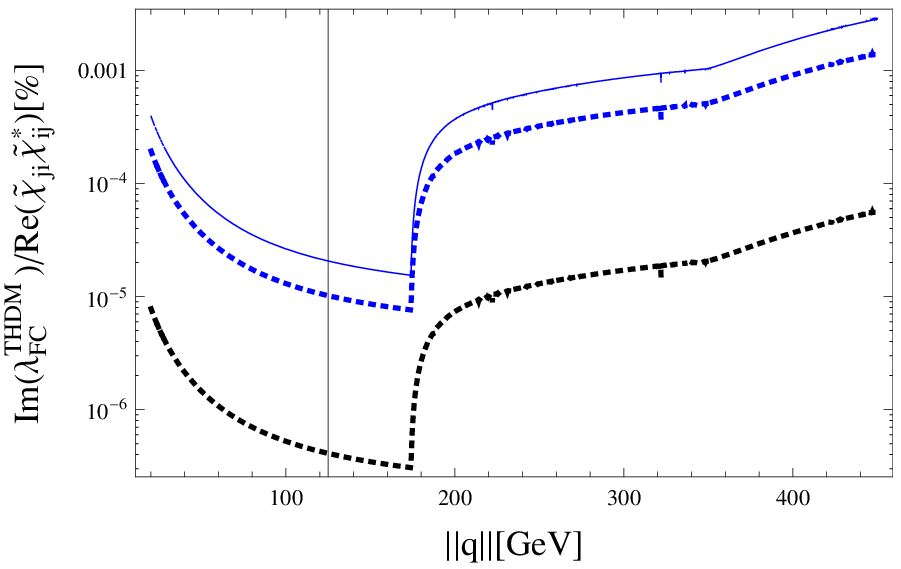}
\caption{The same that in figure \ref{lambda-FV-top-tan1} but for $\tan\beta=10$.}
\label{lambda-FV-top-tan10}
\end{figure}

As mentioned previously, for this case we need to consider only two effective vertices with charm and top quarks, with an arbitrary $\text{Re}(\tilde{\chi}_{ij}^u\tilde{\chi}_{ji}^{u*})$ factor. In Figures \ref{lambda-FV-top-tan1} and \ref{lambda-FV-top-tan10} we depict the numerical behavior of the real (left) and imaginary parts (right) of the $\hat{\lambda}_\text{FC}^\text{THDM-III}$ form factor, as function of the 4-momentum magnitude of the off-shell scalar boson. These figures correspond to $\tan\beta=1$ and $\tan\beta=10$ respectively, we have used $|\chi|=0.20$ (blue line), $|\chi|=0.14$ (blue dotted line) and $|\chi|=0.028$ (blank dotted line). In the first place, we appreciate that the imaginary part is smaller than the real part, although both contributions decrease significantly with smaller values of $\chi$. Unlike the SM contributions, the flavor changing coupling in the THDM type III induces an imaginary part for $||q||<m_t+m_c$. Moreover, the uncoupling behavior is appreciable with an abrupt slope change at $||q||=m_t+m_c$. Finally, for largest values of $\tan\beta$ we found suppressed contributions.


Table \ref{lambda-mh-2HDM-III} shows the predictions induced at the Higgs on-shell scheme, for some values of $\chi$ and $\tan\beta$. As it was expected, the less suppressed contribution would come from smaller values of the $\tan \beta$, and for not very suppressed values of $\chi$. It is important to note that though small, the correction induced by the flavor changing coupling can be larger than the tau lepton or bottom-quark corrections.

\begin{table}[!htb]
\begin{center}
\begin{tabular}{|c|c|c|}
\hline
\backslashbox{$\chi$}{$\tan\beta$}&$\tan\beta=1$ & $\tan\beta=10$  \\
\hline
\hline
$\chi=0.2$ & $1.12\times10^{-3}+i4.08\times10^{-5}$&$1.12\times10^{-3}+i4.08\times10^{-5}$\\
$\chi=0.14$ & $5.54\times10^{-4}+i2.01\times10^{-5}$&$5.54\times10^{-4}+i2.01\times10^{-5}$\\
$\chi=0.028$ & $2.23\times10^{-5}+i8.11\times10^{-7}$&$2.23\times10^{-5}+i8.11\times10^{-7}$\\
\hline
\end{tabular}
\caption{Predictions for $\hat{\lambda}_\text{FC}^\text{THDM-III}/\text{Re}(\tilde{\chi}_{ij}^u\tilde{\chi}_{ji}^{u*})$\% factor for some values of $\chi$ and $\tan\beta$; we consider the Higgs on-shell scheme.}
\label{lambda-mh-2HDM-III}
\end{center}
\end{table}

\subsection{LHM with T-parity framework}
In the LHM with T-parity there are two different contributions of the mirror fermions: one for mirror up quark and another for the mirror heavy neutrino. Consequently, there are two different types of mixing matrices $\kappa_{ii}$, but for our analysis we will consider that such matrices have the same magnitude of order e.g. $\kappa_{i\ell_H}\sim\kappa_{iq_H}\equiv \kappa_{ii}$. Two free parameters of the LHM with T-parity are thus involved in the contribution (\ref{l-LHM+T}): the scale symmetry breaking $f$ and the mixing matrix $\kappa_{ii}$. Constraints on these parameters are discussed in Ref. \cite{Reuter:2013iya}, where it is considered that the mirror quarks are heavier than all the heavy gauge bosons. This corresponds to values of $\kappa_{ii} \gtrsim 0.45$, which makes the decay $q_H\to V_H q$ ($V_H=W_H^\pm,Z_H$) kinematically allowed, while for $\kappa_{ii}\lesssim0.45$ the only kinematically allowed decay of the mirror quark is $q_H\to A_H q$, and finally if $\kappa_{ii}\lesssim0.1$ the mirror quarks are stable. Thus, for $0.1\lesssim\kappa_{ii}\lesssim0.45$ and some results obtained from the 8 TeV run at the LHC, the combined analyses of electroweak precision physics and Higgs precision physics results in a lower bound $f \gtrsim 694$ GeV at 95\% C.L., while $f \gtrsim638$ GeV was obtained from direct searches in $pp\to q_Hq_H$ and $pp\to q_H A_H$ processes at 95\% C.L. Therefore, in the following analysis we will consider $\kappa_{ii}=0.45$, and the region between 500 GeV and 2000 GeV for the scale symmetry breaking. Moreover, for simplicity we consider that the masses of the three mirror families are the same and then the sum over families is replaced by a three factor.

\begin{figure}[!hbt]
\centering
\includegraphics[scale=0.95]{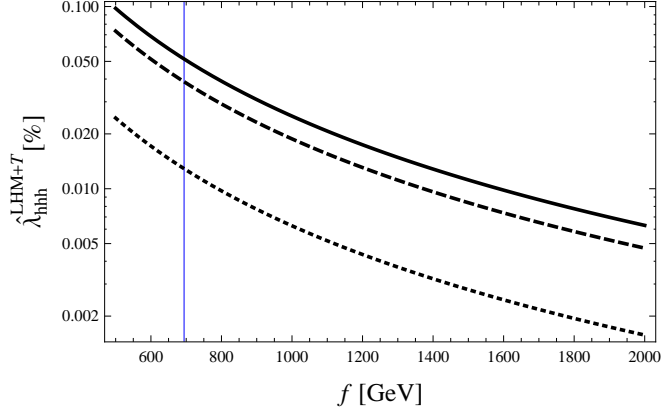}
\caption{Prediction for the $\hat{\lambda}_{hhh}$ form factor in the LHM with T-parity, as a function of the symmetry breaking scale $f$. We have used the Higgs resonance region and $\kappa_{ii}=0.45$ to obtain the heavy mirror up-quark (dashed line), heavy mirror neutrino (dotted line) and the sum of both contributions (black line). The vertical line corresponds to the lower bound $f=694$ GeV.}
\label{lambda-LHM-f}
\end{figure}

\begin{figure}[!hbt]
\centering
\includegraphics[scale=0.95]{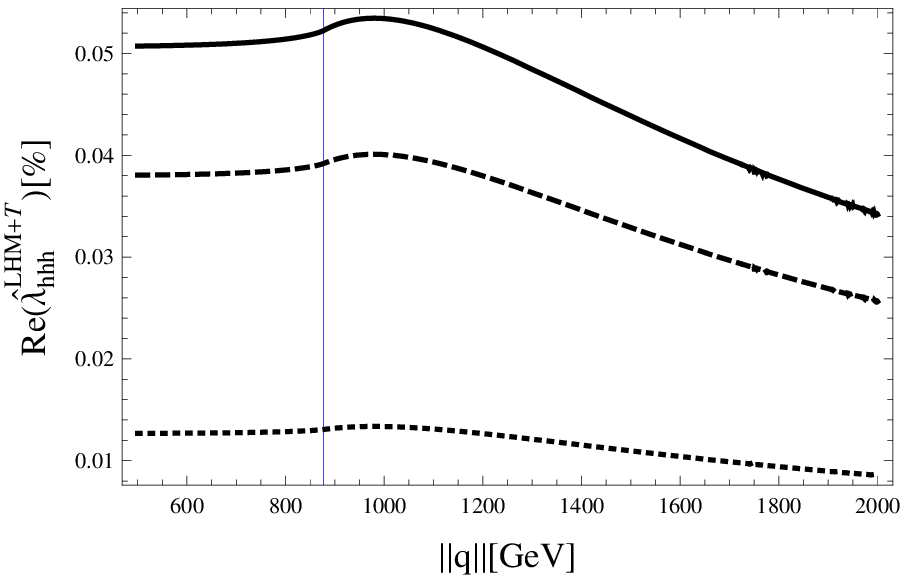}
\includegraphics[scale=0.95]{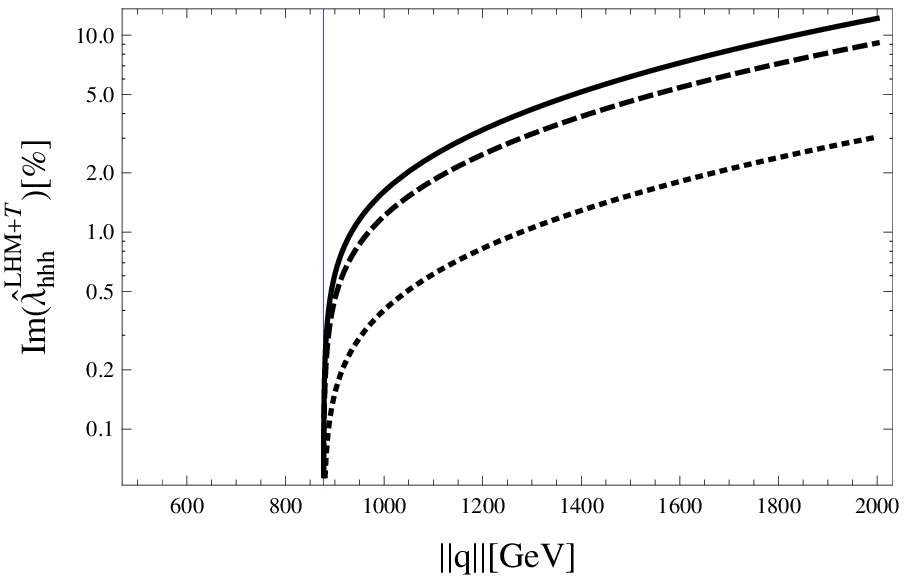}
\caption{Real (left) and imaginary (right) contributions of the $\hat{\lambda}_{hhh}^\text{LHM+T}$ form factor, as function of the 4-momentum magnitude of the off-shell scalar boson. We have used $f=700$ GeV and $\kappa_{ii}=0.45$, to obtain the heavy mirror up-quarks (dashed line), heavy mirror neutrinos (dotted line) and the sum of both contributions (black line). The vertical line corresponds to $||q||=2 m_{f_H}\simeq 877$ GeV.}
\label{lambda-LHM-f700}
\end{figure}

\begin{figure}[!hbt]
\centering
\includegraphics[scale=0.95]{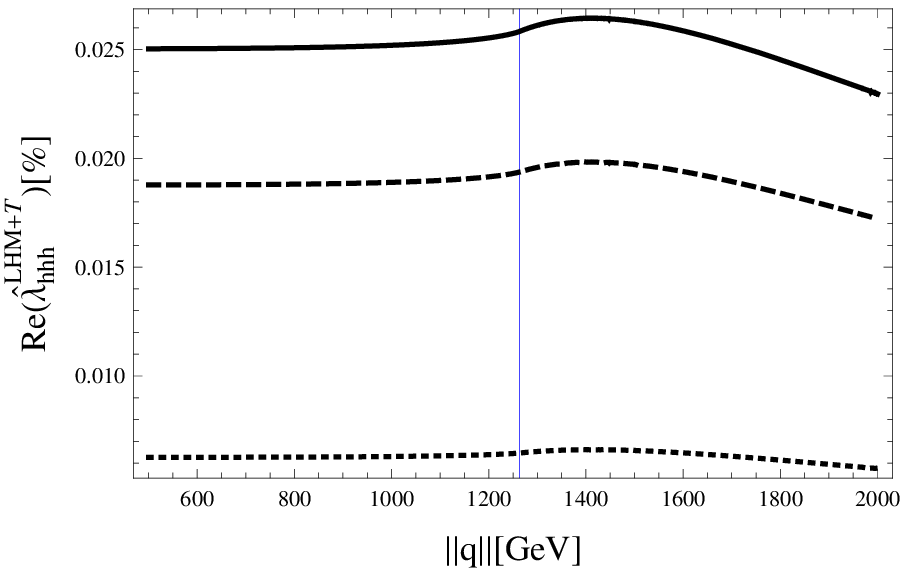}
\includegraphics[scale=0.95]{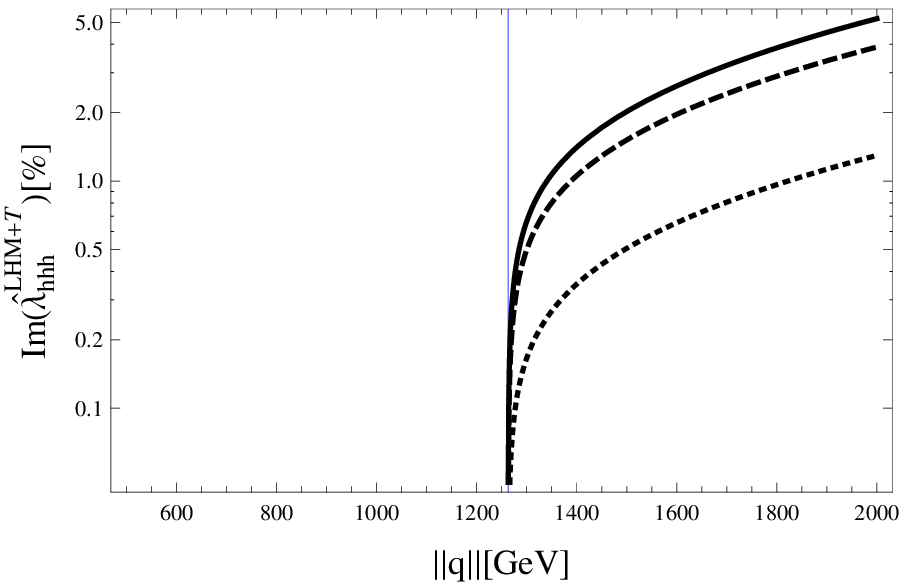}
\caption{The same situation as in figure \ref{lambda-LHM-f700} but for $f=1000$ GeV, where now $2 m_{f_H}\simeq 1263$ GeV.}
\label{lambda-LHM-f1000}
\end{figure}

In the Fig. \ref{lambda-LHM-f} we depict the mirror up quark, mirror neutrino and the sum of both contributions for the $\hat{\lambda}_{hhh}$ form factor, as a  function of the symmetry breaking scale $f$ for $\kappa_{ii}=0.45$ and for the Higgs on-shell scheme. We also include the lower bound $f \gtrsim638$ GeV, which is given by the vertical line in this figure. Mathematically speaking, the difference between the contributions of heavy mirror up-quarks and heavy mirror neutrinos is given by the number of color $N_c$; consequently the mirror up-quark correction is the dominant contribution. We note that these corrections decrease softly with the increasing symmetry breaking scale $f$. Particulary, for the sum of both contributions in $f=500$ GeV we have $\hat{\lambda}_{hhh}^\text{LHM+T}\simeq 0.0977\%$ and for $f=2000$ GeV, we have $\hat{\lambda}_{hhh}^\text{LHM+T}\simeq 0.0062\%$, while that for the lower bound $f=694$ GeV is the $\hat{\lambda}_{hhh}^\text{LHM+T}\simeq 0.0512\%$. It is important to note that there is no imaginary part in the regime of the Higgs on-shell scheme, but for large 4-momentum magnitude of the off-shell scalar boson an imaginary part is induced; Figures \ref{lambda-LHM-f700} and \ref{lambda-LHM-f1000} depict this situation.

For $f=700$ GeV and $\kappa_{ii}=0.45$ we show the real and imaginary parts of the $\hat{\lambda}_{hhh}^\text{LHM+T}$ form factor, as function of the 4-momentum magnitude of the off-shell scalar boson in Figure \ref{lambda-LHM-f700}. The numerical behavior is similar to the fermionic SM contribution (see Fig. \ref{lambda-sm-top}), but in a smaller scale. There is a small maximum in the real part after $2 m_{f_H}\simeq 877$ GeV, where also the imaginary part is induced. Moreover, the imaginary part increases with the 4-momentum magnitude of the off-shell scalar boson, such as the top quark in the SM, which is considerably reduced by the contributions of heavy gauge bosons and self-coupling contributions. In particular, we have $\text{Re}(\hat{\lambda}_{hhh}^\text{LHM+T})\simeq 0.0507\%$ and $\text{Re}(\hat{\lambda}_{hhh}^\text{LHM+T})\simeq 0.0341\%$ for the real part of the sum of both contributions, in $||q||=500$ GeV and $||q||=2000$ GeV, respectively.

If the symmetry breaking scale is increased to $f=1000$ GeV, the corrections decrease softly and we have the same behavior than the previous case, but now with $2 m_{f_H}\simeq 1263$ GeV. In this case we have used the same values of $||q||$ than the previous case, to obtain $\text{Re}(\hat{\lambda}_{hhh}^\text{LHM+T})\simeq 0.0250\%$ and $\text{Re}(\hat{\lambda}_{hhh}^\text{LHM+T})\simeq 0.0229\%$ for the sum of both contributions, respectively. Thus, we can appreciate that the correction to the $\hat{\lambda}_{hhh}$ form factor in LHM with T-parity is not very sensitive to higher values of the symmetry breaking scale $f$, and of the 4-momentum magnitude of the off-shell scalar boson $||q||$.

\section{Conclusions}
We have performed a detailed calculation of the one-loop radiative corrections to the triple Higgs self-coupling ($\lambda_\text{loop}$) in the framework of the SM, the THDM type III and the LHM with T parity. Since the results are UV-divergent, we analyzed the respective renormalization scheme and we studied the correspondence of our results with results previously reported. Then, we found that the UV-divergence terms vanished with the respective counterterms, and further they have no contribution to the final result $\hat{\lambda}_{hhh}$. Appendix A includes details of the renormalization scheme used in our calculation with a explicit treatment of the divergent contributions.   Numerically speaking, we found that the form factor at one-loop level, with at least one off-shell mass scalar boson, may contain real and imaginary parts. In the SM, the top quark loop dominates the radiative correction, and we were able to reproduce the result obtained in previous studies $9.14049$\% without requiring any approximation for the 4-momentum magnitude of the off-mass shell scalar boson. We also found that in the Higgs on-shell scheme, the virtual effect of the $hhh$ self-coupling induces a radiative correction $\lambda_h^\text{SM} \simeq 1.83974\%$, which in turn modifies the total radiative SM correction $\hat{\lambda}_{hhh}^\text{SM}=\hat{\lambda}_\text{top}^\text{SM}+\hat{\lambda}_{W+Z}^\text{SM}+\lambda_h^\text{SM}\simeq 11.0528\%$.

Our results for the top-quark correction to the $hhh$ form factor shows that an imaginary part is generated when the momentum magnitude of the off-mass shell Higgs boson leg is larger than the Higgs boson mass. This imaginary part is larger than the respective real part, but we showed that virtual effects of the $hhh$ self-coupling and the gauge boson contributions ($V=Z^0$, $W^\pm$) reduce the magnitude of the top-quark imaginary part, without further consequences to the unitary constraints. However, the real and imaginary parts to the one-loop corrected $h^*hh$ form factor should be considered in a complete calculation of the cross section for the process $gg \to hh$ for energies beyond the on-mass shell threshold. Kanemura et al. \cite{Kanemura:2016lkz} have already performed a detailed calculation of this cross section on-mass shell in the SM and some extensions of the SM. They found that sizable effects could be generated for the SM result if the one-loop corrected $hhh$ vertex has large  deviations. A complete study of the $gg \to hh$ process for energies beyond the on-mass shell threshold will be left for a future work.

On the other hand, we also computed in the THDM type III the radiative correction induced by a FCNC of the type $h^0tc$. Our results show that this new fermionic contribution is also small but larger than the lepton and bottom-quark corrections. Finally, we obtained that the new degrees of freedom associated to the THDM type III and the LHM  with T-parity are rather smaller, and for the last framework they are not very sensitive to the value of the symmetry breaking scale $f$.

\section*{Acknowledgement}
We acknowledge support from CONACyT and SNI (Mexico).

\appendix

\section{Renormalization of self-coupling $hhh$ at one-loop level}
In this appendix we discuss the  treatment to cancel the UV divergences of the self-coupling $hhh$. Since the main contributions is induced by heavy fermion, we focus mainly in these results. The respective counterterms of the equation(\ref{deltalambda}) are introduced according to following redefinitions:

\begin{eqnarray}
e&\to&(1+\delta Z_e)e, \nonumber\\
s_W&\to&s_W+\delta s_W, \nonumber\\
m_h^2&\to&m_h^2+\delta m_h^2, \nonumber\\
m_W^2&\to&m_W^2+\delta m_W^2, \nonumber\\
h^0&\to&(1+\frac{1}{2}\delta Z_h)h^0. \nonumber\\
\end{eqnarray}
The Higgs tadpole contribution is zero at one-loop level and thus $0=T +\delta t$, where $T$ is the Higgs-boson one-point function. The other counterterms involve two-point functions of the photon $\Sigma_{T}^{AA}$, the Z-photon mixing $\Sigma_{T}^{AZ}$, the gauge bosons $\Sigma_{T}^{VV}$ and the Higgs boson $\Sigma_{T}^{h}$, in the following form \cite{Bohm:1986rj,Denner:1991kt}:

\begin{eqnarray}
\delta Z_{e}&=&\frac{1}{2}\frac{\partial \Sigma_{T}^{AA}(q^{2})}{\partial q^{2}}\bigg{\vert}_{q^{2}=0}-\frac{s_{W}}{c_{W}}\frac{\partial \Sigma_{T}^{AZ}(0)}{m_{Z}^{2}}\\
\frac{\delta s_{W}}{s_{W}}&=&-\frac{1}{2}\frac{c_{W}^{2}}{s_{W}^{2}} \widetilde{\mathrm{Re}}\left(\frac{\Sigma_{T}^{WW}(m_{W}^{2})}{m_{W}^{2}}-\frac{\Sigma_{T}^{ZZ}(m_{Z}^{2})}{m_{Z}^{2}} \right)
\end{eqnarray}

\begin{equation}
\delta Z_{h}=-\mathrm{Re}\frac{\partial \Sigma^{h}(q^{2})}{\partial q^{2}}\bigg{\arrowvert}_{q^{2}=m_{h}^{2}}, \qquad \delta m_{h}^{2}=\mathrm{Re} \Sigma ^{h}(m_{h}^{2}), \qquad \delta m_{W}^{2}=\widetilde{\mathrm{Re}}\Sigma_{T}^{WW}(m_{W}^{2}).
\end{equation}

The complete expressions of all these two-point functions are included in the Eqs. (B1)-(B5) of the reference \cite{Denner:1991kt}, where the full contributions of each one are presented and they involve the Passarino-Veltman functions $A_0(m^2)$ and $B_0(k^2,m_1^2,m_2^2)$. In our case, the top-quark counterterm $\delta\lambda_t^\text{SM}$ for $\lambda_f^\text{SM}$ given in Eq.(11) is constructed out  of the respective Higgs-boson one-point function and the Higgs-boson contributions to $\Sigma_{T}^{AA}$, $\Sigma_{T}^{AZ}$, $\Sigma_{T}^{VV}$ and $\Sigma^{h}$. In an analogous way, we constructed the $\delta\lambda_V^\text{SM}$ counterterm associated to the gauge-boson contributions to $\lambda_{V}^\text{SM}$ given in Eqs.(14) and (15). On the other hand, since only the bubble loop induced by the Higgs-boson self coupling is UV-divergent (\ref{lambda_hII}), the respective  $\delta\lambda_{h_{II}}^\text{SM}$ receive the contributions from $\Sigma_{T}^{VV}$, $\Sigma^{h}$ and the Higgs-boson one-point induced by the $hhh$ self-coupling.

The above renormalization scheme has been used satisfactorily in the radiative corrections to the vertices $hhh$ and $hZZ$ at one-loop level \cite{Arhrib:2015hoa,He:2016sqr,Baglio:2016ijw}. However, in our case we have also contributions coming from flavor-changing neutral couplings involving the top-quark in the THDM and mirror fermions of the LHM with T-parity. In order to include these two new corrections in our renormalization scheme, first we need to calculate the two-point functions induced by the flavor-changing neutral couplings:

\begin{equation}
\Sigma_\text{FC}^{h}(q^2)=\frac{N_c g^2m_im_j(\xi_H^u-\xi_h^u\cot \beta)^2}{2^8\pi^2m_W^2(1+\cot^2\beta)}\text{Re}(\tilde{\chi}_{ij}^u\tilde{\chi}_{ji}^{u*})\Big[A_0(m_i)+A_0(m_j)+\big[(m_i+m_j)^2-q^2\big]B_0(q^2,m_i^2,m_j^2)  \Big].
\end{equation}

With appropriate changes of parameters and masses, the above equation reproduce the fermionic contribution for the SM Higgs two-point function $\Sigma_{T}^{h}$ given by the Eq. (B.5) of the reference \cite{Denner:1991kt}. For the purpose of the present calculation, it is convenient to express the Higgs two-point function in terms of Feynman parameters as follows,

\begin{eqnarray}
\Sigma_{T}^{h}(q^2)&=&\frac{N_c g^2m_im_j(\xi_H^u-\xi_h^u\cot \beta)^2}{2^83\pi^2m_W^2(1+\cot^2\beta)}\text{Re}(\tilde{\chi}_{ij}^u\tilde{\chi}_{ji}^{u*})\bigg[3(2\Delta+1)(m_i^2+m_j^2)+6m_i^2m_j^2\Delta-q^2(3\Delta+1)
                                        \nonumber\\
{}&{}&-6\int_{x=0}^1\text{d}x\big[x(x-1)q^2+m_im_j+2[m_jx+(q^2x-m_i^2)(x-1)]\big]
                                        \nonumber\\
{}&{}&\log\Big(\frac{m_jx+(q^2x-m_i^2)(x-1)}{\Lambda}\Big) +(i\rightleftarrows j)\bigg].
\end{eqnarray}
Similarly, with appropriate changes, we obtain the respective SM contribution in an exact form,
\begin{eqnarray}
\Sigma_\text{SM}^{h}(q^2)&=&\frac{N_cm_f^2\alpha}{8\pi s_W^2m_W^2}\bigg[2m_f^2(3\Delta+5)-q^2(\Delta+2)+(q^2-6m_f^2)\log\Big(\frac{m_f^2}{\Lambda}\Big)
                                        \nonumber\\
{}&{}&-\frac{2}{q}(4m_f^2-q^2)^{3/2}\tan^{-1}\bigg(\frac{q}{\sqrt{4m_f^2-q^2}}\bigg)\bigg]\label{S-smf}.
\end{eqnarray}
On the other hand, the Higgs-boson one-point function can not be induced by the flavor-changing neutral couplings but we need to include it in our renormalization scheme and we used a linear combination of one-point functions,

\begin{equation}
T^{h}(m_i^2,m_j^2)=\frac{T(m_i^2)+T(m_j^2)}{2},
\end{equation}
where
\begin{eqnarray}
T(m_k^2)=\frac{N_c g m_k^4}{8\pi^2 m_W}\tilde{h}_{kk}\bigg[1+\Delta-\log\bigg(\frac{m_k^2}{\Lambda^2}\bigg)\bigg].\label{Tad-f}
\end{eqnarray}
It is important to mention that we obtain the SM  results for $\Delta \Gamma_H^{(1,2)}$ given in Eq. (29) of \cite{Hollik:2001px}, with $q^2/m_t^2\simeq 0$ in (\ref{S-smf}) and $\tilde{h}_{kk}=1$ in (\ref{Tad-f}). Therefore, our renormalization scheme is just a generalization of the usual procedure used for the renormalization of the SM fermionic radiative corrections to the $hhh$ vertex. In the case of the flavor-conserving neutral correction, the renormalization process is basically the same used in the SM, except for the proportionality factor $\tilde{h}_{ii}^2$. The same situation is obtained for the mirror fermion contribution in the LHM with T-parity. In this case the masses of these fermions depend on three parameters (Eq. (7)): the diagonal element $\kappa_{ii}$, the energy scale $f$ and the vacuum expectation value $v$. However, it is convenient to take each mirror fermion mass as a free parameter. In this way, we get the same mathematical structure for the mirror fermion contribution, Eq. (\ref{l-SM2}), than the SM result given in Eq. (\ref{l-LHM+T}). The only difference in both contributions is associated to the respective Higgs boson couplings to fermions and without losing generality, for the correspondence $(g,m_f,m_W)\to(k_{ii},v,f)$ can be applied. As a consequence, the SM fermionic contributions have the same renormalization scheme in the LHM with T-parity.

\section{Feynman parameters}

In the calculation of Eq. (\ref{l-eff2}), we needed six different permutations for the 4-momentum of the external Higgs bosons and we used the following Feynman  parameters,

\begin{eqnarray}
M_1^2&=& x^2+y^2-s_j^2(x+y-1)-(s_q^2-2)xy+(s_i^2-1)(x+y),
                                        \nonumber\\
M_2^2&=&x^2+s_i^2y-s_j^2(x+y-1)+s_q^2(y-1)y+x(s_i^2+s_q^2y-1),
                                        \nonumber\\
M_3^2&=&y(y-1)+s_i^2(x+y)+(s_q^2x-s_j^2)(x+y-1)\label{Mk}.
\end{eqnarray}
In the flavor conserving case $i=j$ and the $M_k^2$ functions reproduce the respective functions $\mu_k^2$ for the SM fermionic contributions given in Eq. (\ref{l-SM2}).

\bibliography{hhh-v1}{}

\end{document}